\documentstyle[12pt,epsf,axodraw]{article}
\def\gsim{\mathrel{\rlap {\raise.5ex\hbox{$ > $}}
{\lower.5ex\hbox{$\sim$}}}}
\def\lsim{\mathrel{\rlap {\raise.5ex\hbox{$ < $}}
{\lower.5ex\hbox{$\sim$}}}}

\newcommand{\be}{\begin{equation}}
\newcommand{\ee}{\end{equation}}
\newcommand{\bea}{\begin{eqnarray}}
\newcommand{\nn}{\nonumber}
\newcommand{\eea}{\end{eqnarray}}
\newcommand{\nd}[1]{/\hspace{-0.6em} #1}

\def\gappeq{\mathrel{\rlap {\raise.5ex\hbox{$>$}}
{\lower.5ex\hbox{$\sim$}}}}
\def\lappeq{\mathrel{\rlap{\raise.5ex\hbox{$<$}}
{\lower.5ex\hbox{$\sim$}}}}

\begin{document}
 
\begin{titlepage}
\begin{flushright}
CERN-TH/99-94\\
hep-th/9904046 \\
\end{flushright}

\begin{centering}
\vspace{.1in}

{\large {\bf Non-linear Dynamics in ${\rm QED}_3$
and }} \\[0.4cm]
{\large {\bf  Non-trivial Infrared Structure }}
\\
\vspace{.2in}

{\bf  N.E. Mavromatos$^{a,b}$
and J. Papavassiliou$^{b}$} \\
[0.4cm]
$^{a}$ Department of Physics
(Theoretical Physics), University of Oxford, \\1 Keble Road,
Oxford OX1 3NP, U.K.  \\[0.4cm]
$^b$ CERN Theory Division, Geneva 23 CH-1211, Switzerland. \\[0.3cm]
\vspace{0.5in}
 
{\bf Abstract} \\
\vspace{0.1in}
\end{centering}
{\small In this work we consider a coupled 
system of Schwinger-Dyson equations
for self-energy and vertex functions 
in $QED_3$.
Using the concept of a semi-amputated vertex
function, we manage to decouple the vertex equation
and transform it in the infrared into a non-linear differential equation 
of Emden-Fowler type. 
Its solution suggests the following picture:
in the absence of infrared cut-offs there is only a trivial
infrared  
fixed-point structure in the theory. 
However, the presence of masses, for either fermions or photons, 
changes the situation drastically, leading to a 
mass-dependent non-trivial infrared fixed point. 
In this picture a dynamical mass for the fermions is found 
to be generated consistently. 
The non-linearity of the equations  
gives rise to 
highly non-trivial 
constraints among the mass and effective (`running')  
gauge coupling, which impose lower and upper bounds on the latter
for dynamical mass generation to occur.  
Possible implications of this to the theory of 
high-temperature superconductivity are briefly discussed.}

\vspace{3cm}
\begin{flushleft}
CERN-TH/99-94\\
April 1999 \\
\end{flushleft}

\end{titlepage} 

\newpage

\section{Introduction}

Three-dimensional Quantum Electrodynamics ($QED_3$),
with an even number of fermion flavours, 
apart from serving as a toy model for studying chiral-symmetry 
breaking and confinement, also constitutes a
physically interesting theory {\it per se}, in view of its possible 
applications in modelling novel (high-temperature) 
superconductors~\cite{dor}-\cite{csm}. 

Chiral symmetry breaking or, equivalently, dynamical 
mass generation for fermions, in even-flavour
$QED_3$ has still many unresolved issues.
One of those is the existence of a (dimensionless) critical coupling,
above which dynamical mass generation for the fermions 
occurs~\cite{app}. 
In the context of large-$N$ treatment, which at present constitutes
the only well-studied approach, the r\^ole of the dimensionless
coupling~\footnote{In three dimensions the coupling $e^2$ has dimensions of mass. 
One can still define, however, dimensionless couplings by 
dividing with a dynamically generated scale, which in the large $N$-treatment
arises by demanding that~\cite{app} 
$e^2 N =8 \alpha$, with 
the scale $\alpha$ kept fixed as $N \rightarrow \infty$. 
The dimensionless
coupling is then defined as the ratio $e^2/8\alpha = 1/N$.}
is played by the inverse of the fermion flavour number
$N$. 
The issue of the existence of a critical coupling in $QED_3$ is a delicate 
one~\cite{penn}. Many of 
the original approximations~\cite{app} leading to its existence 
have  been questioned, in particular 
the fact that wave-function 
renormalization effects have not been properly 
accounted for.  
Recently, however, the incorporation of such effects,
still within the large -$N$ context, 
appears to 
corroborate~\cite{amm}-\cite{maris} the qualitative picture 
advocated in \cite{app}. 
In addition, the latter is also supported 
by 
lattice simulations~\cite{kogut}. 

Nonetheless, the situation is far from being conclusive. 
The fact that the critical $N$, below which dynamical mass 
generation 
occurs, 
is found to be of order $3$ (in a four-component notation for the fermions) 
provides motivation for searches beyond the large-$N$ treatment. 
Moreover, up to now the gauge coupling in the infrared has been treated
as an arbitrary parameter, whose size has not been restricted
by an additional  dynamical constraint. 
At present we are lacking 
a self-consistent treatment of the 
dynamical Schwinger-Dyson (SD) equations 
involving the vertex function on an equal footing 
with the self-energy and gap functions.  
In all the approaches so far, at least within the large $N$ treatment
that we are aware of, 
one invokes a specific Ansatz for the vertex, by the sole requirement   
of satisfying some truncated form of the Ward-Takahashi identity 
stemming from gauge invariance~\cite{kondo}-\cite{am}. 
The lack of dynamical information for the coupling 
poses problems; for instance, 
its size  
in the infrared is treated as an arbitrary parameter, 
being assumed to merely exceed a 
critical value, if one wishes to 
trigger chiral symmetry breaking.

Another point, related to the above, which is 
already familiar from studies in the case of 
four-dimensional non-Abelian gauge theories, 
is whether chiral-symmetry breaking is associated with confinement 
of charges~\cite{conf}. 
This issue acquires physical importance 
in view of 
the condensed-matter applications. In particular, 
it may shed more 
light in the dynamics of spin-charge separation, by analogy with the 
physics of strong interactions~\cite{laughlin}. 

In this work we shall not deal with 
issues of confinement, which exists in $QED_3$
despite its abelian nature. Instead, we shall attempt a novel 
approach to chiral symmetry breaking, 
independently of a large $N$ treatment, by 
studying the {\it coupled} fermion and photon self-energies and vertex 
SD equations in the context of a 
method first introduced for the case of four-dimensional 
$QCD$~\cite{pc}. 
The novel ingredient is that we concentrate on the semi-amputated 
vertex, defined in section 2, which is the correct gauge-invariant quantity
to determine a physically meaningful ``running'' coupling 
(`effective charge').
The fact that $QED_3$ is superrenormalizable in the ultraviolet does not 
preclude the possibility of defining such a quantity, having non-trivial 
structure in the infrared. 
As we show in sections 2 and 3, the existence of a non-trivial infrared fixed
point is a property {\it only} of the infrared-regularized
theory, as conjectured in \cite{am}. In other words, in the absence of 
fermion (or photon) masses, the infrared singularities of the 
vertex equation will force the effective charge to vanish at zero momentum 
transfer, thereby excluding the possibility of an interesting
infrared behaviour. From a condensed-matter application point of view this 
would correspond to what is usually called the Landau-Fermi liquid theory 
(trivial infrared fixed point)~\cite{shankar,am}. 
In the presence of masses, and in particular fermion masses, 
we show in 
section 3 that there is a non-trivial infrared fixed point structure,
stemming from the fact that the effective charge obtained as a self-consistent 
solution of the {\it non-linear} vertex SD equation, is driven to a 
finite positive value, which can be large enough to trigger dynamical 
generation of a fermion  mass. This implies that 
the phenomenon of chiral symmetry breaking is intimately 
associated with deviations from the trivial infrared fixed-point structure.

We should stress that, as a result of the non-linearity of the 
vertex equation, there are delicate constraints 
between the fermion  mass and the effective charge, 
which are responsible for the appearance of regions of the latter
for which dynamical generations occurs. At present, these restrictions 
appear as a consequence of mathematical self-consistency of the 
truncated equations. It is not clear to us, whether the upper bounds 
on the effective charge, imposed by the present cubic approximations 
for the vertex corrections, 
will survive the inclusion of higher orders. 
In contrast, we believe that the lower bounds
will survive such a treatment, thereby indicating 
the existence of a critical 
coupling above which dynamical mass generation will occur. 
This is physically appealing, given that one would not expect 
a weakly-coupled theory to be capable of breaking 
dynamically chiral symmetry.
 
The layout of this article is as follows: in section 2
we set up the SD equations that we wish to study, 
and discuss in detail the approximations employed. 
We demonstrate that,
under certain assumptions to be justified
retrospectively in section 4, 
the equation 
for the semi-amputated vertex 
decouples from the rest, 
and hence can be solved separately.  
Moreover, we establish the absence of a non-trivial infrared fixed point
rigorously (within the cubic approximation for the vertex), 
by casting the SD equation for the 
effective charge in a form of a non-linear differential 
equation, known as Emden-Fowler equation~\cite{emden,kamke}. 
In section 3 we study the equation for the vertex 
in the presence of a fermion mass, acting as an infrared regulator. 
We derive the appropriate non-linear differential  equation describing 
the infrared behaviour of the running coupling, and solve it to demonstrate
the existence of a non-trivial infrared fixed point. 
The so-derived running charge is a monotonically decreasing function 
of the momentum, tending asymptotically to a constant positive value 
in the ultraviolet.  
In section 4, we solve the equations for the photon and fermion 
self-energies, which in our approximations decouple from each other
and depend only on the semi-amputated vertex function. We then  
verify the self-consistency of the approach. 
In section 5 we examine the self-consistency 
of the dynamical generation of the fermion mass, by 
solving the appropriate SD equation
upon substituting the solution for the 
semi-amputated vertex found in previous sections. 
The self-consistency of the approach restricts the 
allowed regions of the effective charge, implying the existence
of a lower bound (critical coupling) but also of an upper one. 
In section 6, we examine an alternative type of infrared cut-off, namely
that of a (bare) covariant photon mass term. This case also 
exhibits a non-trivial infrared fixed-point structure but, 
in contrast to the monotonic decrease of the effective charge in the 
case of fermion masses, here the coupling initially increases 
in the infrared, then displays a local maximum, 
and eventually decreases, tending asymptotically to a constant value 
in the ultraviolet. Some possible applications of this 
behaviour, inspired by condensed-matter physics, are briefly discussed.
Finally, in section 7 we present our conclusions and outlook.

\medskip

\setcounter{equation}{0}
\section{The SD equation for the semi-amputated vertex}

In this section we will first set up the SD equations
for the photon and electron self-energies,
 and the photon-electron vertex;
then we will define the semi-amputated vertex and derive its corresponding
 SD equation. As we will explain, the latter 
governs the behaviour of the effective coupling in
the infra-red.
The derivation of the SD equations for the 
photon propagator $\Delta_{\mu\nu}$, the electron
propagator $S_{F}$, and the photon-electron vertex
$\Gamma_{\mu}$ proceeds following standard methods
\cite{cjt,cm} (see figures \ref{fig1a},\ref{fig1b},\ref{fig1c}).

The full photon propagator $\Delta_{\mu\nu}$, 
its inverse $\Delta_{\mu\nu}^{-1}$, and the full  vacuum polarisation
$\Pi_{\mu\nu}$ in Euclidean space are related by
\bea
\Delta_{\mu\nu}^{-1}(q) &=& 
\Delta^{-1}_{0\mu\nu}(q) + \Pi_{\mu\nu}(q) \nonumber\\
{\Delta}^{-1}_{0\mu\nu}(q) &=& 
q^2 \delta_{\mu\nu} - (1-\frac{1}{\xi})q_{\mu}q_{\nu}\nonumber\\
\Pi_{\mu\nu}(q) &=& (q^2~\delta_{\mu\nu} - q_{\mu}q_{\nu})\Pi(q) \nonumber\\
\Delta_{\mu\nu}(q) &=& 
(\delta_{\mu\nu} - q_{\mu}q_{\nu}/q^2)[q^2 + q^2~\Pi(q)]^{-1}
+ \xi q_{\mu}q_{\nu}/q^4
\label{vacpol}
\eea
where $\xi$  is the  gauge fixing  parameter  (in  covariant gauges).
The corresponding SD  equation reads
\be 
\Delta_{\mu\nu}^{-1}(q)=
\Delta^{-1}_{0\mu\nu}(q)   +  e^2  \int  \frac{d^3k}{(2\pi)^3}    
{\rm Tr}[\Gamma_{\mu} S_{F}\Gamma_{\nu}S_{F}] ~+\dots   
\label{SD2}
\ee 
The  SD equation for the electron  propagator   $S_{F}$ is given by  
\be
S_{F}^{-1}(p)= -i\nd{p} - e^2 \int  \frac{d^3k}{(2\pi)^3}
\Gamma_{\mu}  S_{F}\Gamma_{\nu}\Delta^{\mu\nu}    ~+  \dots 
\label{SD1}
\ee 
Finally,   
the SD equation for the photon-electron vertex $\Gamma_{\mu}$ has the form
\bea   
\Gamma_{\mu}(p_1,p_2,p_3) &=& 
\gamma_{\mu} -e^2  \int \frac{d^3k}{(2\pi)^3}                          
\Gamma_{\alpha} S_{F}(k-p_1)\Gamma_{\mu}S_{F}(k)
\Gamma_{\beta}\Delta^{\alpha\beta}(k+p_2) \nonumber\\
&& +\dots 
\label{SD3}
\eea 
with  $p_1+p_2+p_3=0$.  
The ellipses on the right-hand side
of (\ref{SD2})-(\ref{SD3}),
denote the infinite set of terms containing the two-particle
irreducible
four-point function \cite{cjt,cm}.
Although we are not working in the context of a large $N$ analysis, 
we note
that the above truncation is compatible with working to leading 
order in resummed $1/N$ expansion.

\begin{figure}[hbt]
\begin{center}
\begin{picture}(300,50)
\Text(5,30)[]{$\Big($}
\Photon(10,30)(35,30){3}{2.5}
\Vertex(40,30){5}
\Photon(45,30)(70,30){3}{2.5}
\Text(80,31)[]{$\Big)^{-\!1}$}
\Text(95,30)[]{$=$}
\Text(110,30)[]{$\Big($}
\Photon(115,30)(175,30){3}{5.5}
\Text(185,31)[]{$\Big)^{-\!1}$}
\Text(200,30)[]{$+$}
\Photon(215,30)(240,30){3}{2.5}
\Photon(270,30)(295,30){3}{2.5}
\GOval(255,43)(6,8)(0){0}
\GOval(255,17)(6,8)(0){0}
\CCirc(255,30){14}{White}{White}
\CArc(255,30)(15,0,360)
\GCirc(240,30){4}{0.2}
\GCirc(270,30){4}{0.2}
\end{picture}
\caption{Schematic form of the 
  SD equation for the gauge field propagator in resummed
  perturbation theory. The blobs on the right-hand-side indicate the
  full (non-perturbative) vertex corrections.}
\label{fig1a}
\end{center}
\end{figure}

\begin{figure}[hbt]
\begin{center}
\begin{picture}(300,50)(0,0)
\Text(5,30)[]{$\Big($}
\Line(10,30)(70,30)
\Text(80,31)[]{$\Big)^{-\! 1}$}
\GOval(40,30)(6,8)(0){0}
\CBox(32,23)(48,29){White}{White}
\Text(95,30)[]{$=$}
\Text(110,30)[]{$\Big($}
\Line(115,30)(175,30)
\Text(185,31)[]{$\Big)^{-\! 1}$}
\Text(200,30)[]{$-$}
\Line(215,30)(295,30)
\GOval(255,30)(6,8)(0){0}
\CBox(247,23)(263,29){White}{White}
\PhotonArc(255,30)(25,9,79){-3}{2.5}
\PhotonArc(255.4,30)(25,101,171){-3}{2.5}
\Vertex(255,55){5}
\GCirc(230,30){4}{0.2}
\GCirc(280,30){4}{0.2}
\end{picture}
\caption{Schematic form of the SD equation for the full
fermion propagator.  Blobs indicate full non-perturbative
quantities. Notice the full vertex appears on both 
ends of the internal photon line.
\label{fig1b}}
\end{center}
\end{figure}

\begin{figure}[hbt]
\begin{center}
\begin{picture}(330,65)(0,0)
\ArrowLine(0,30)(30,30)
\GCirc(34,30){4}{0.2}
\ArrowLine(38,30)(68,30)
\Photon(34,64)(34,34){3}{3}
\Text(83,30)[]{$=$}
\ArrowLine(98,30)(128,30)
\GCirc(129,30){1}{0.2}
\ArrowLine(130,30)(160,30)
\Photon(129,64)(129,31){3}{3}
\Text(175,30)[]{$+$}
\ArrowLine(190,30)(205,30)
\GCirc(209,30){4}{0.2}
\Line(213,30)(277,30)
\GCirc(245,30){4}{0.2}
\GCirc(281,30){4}{0.2}
\ArrowLine(285,30)(300,30)
\GOval(227,30)(6,8)(0){0}
\GOval(263,30)(6,8)(0){0}
\CBox(218,23)(236,29){White}{White}
\CBox(254,23)(272,29){White}{White}
\Photon(245,64)(245,34){3}{3}
\PhotonArc(245,64.7)(50,228,266){3}{3}
\PhotonArc(245,64.7)(50,274,312){-3}{3}
\Vertex(245,14.7){5}
\Text(315,30)[l]{$+\quad \cdots$}
\end{picture}
\caption{The SD equation for the vertex $\Gamma_{\mu}$ 
\label{fig1c}}
\end{center}
\end{figure}
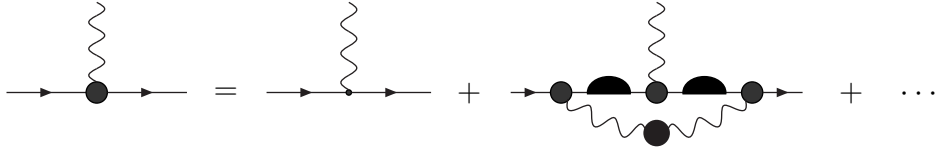

We next define the scalar quantities 
$A$, $B$ and ${\cal G}$ as follows:
\be
S_{F}(k)= \frac{1}{A(k)~\nd{k}}
\label{pertfermion}
\ee
\be   
\Delta _{\mu\nu}(k) \equiv \frac{\delta_{\mu\nu}}{B(k)k^2}
\label{photon}  
\ee
and
\be
     \Gamma _\mu (p_1,p_2,p_3)={\cal G}(p_1,p_2,p_3) \gamma _\mu ~. 
\label{vertexgen}
\ee 
The quantity $B$ is related to $\Pi$ defined in (\ref{vacpol})  
by $B(q)= 1 + \Pi(q)$. 
The definition in (\ref{photon}) implies 
that the longitudinal pieces of the photon propagator will
be discarded in what follows. 
Of course, there are
no rigorous field-theoretic arguments justifying their omission or
inclusion. The correct treatment of such terms  necessitates 
a formalism which would allow for the self-consistent 
truncation of the SD series in a manifestly gauge-invariant way;
unfortunately, no such formalism exists to date.
The standard lore when writing down SD equations is to use in (\ref{SD1})
the form of $\Delta_{\mu\nu}$ given in (\ref{vacpol}), setting $\xi=0$ 
(Landau gauge).
While in $QED_4$ this choice renders the vertex corrections  
unimportant in the ultraviolet, it appears to be less compelling in the context
of the superrenormalizable $QED_3$. In addition, it is known that, while the
conventionally defined fermion self-energy and photon-fermion vertex
depend explicitly on the gauge-fixing parameter $\xi$, it is possible
to construct --at least at one-loop -- a $\xi$-independent  
fermion self-energy and vertex, by resorting to the diagrammatic 
rearrangement of the $S$-matrix known as the pinch technique \cite{PT}.
It turns out that the 
 $\xi$-independent fermion self-energy and vertex
so constructed {\it coincide}
with their conventional counterparts, if we choose for the 
latter the special value $\xi=1$ (Feynmann-t' Hooft gauge)~\cite{PT2}. 
Furthermore, 
as  has  been formally  shown    in
\cite{Stagnant}, {\it all} longitudinal pieces 
appearing in  (\ref{vacpol}) 
vanish from  {\it physical}
observables,    such   as $S$-matrix   elements,  to    all  orders in
perturbation theory.  Thus, one is  led to a  generalized form of
the  Feynman-t' Hooft gauge,  known as the  ``stagnant gauge'',  where only the
$\delta_{\mu\nu}$ part of the vacuum polarization
contributes, to {\it all} orders in perturbation theory.
This gauge will be adopted throughout the present article.  

Following \cite{pc} and \cite{cm} 
we define the semi-amputated vertex ${\hat G}$ as 
\be 
     {\hat G}(p_1,p_2,p_3) \equiv Z(p_1,p_2,p_3) {\cal G}(p_1,p_2,p_3)
\ee
\label{amp}
with 
\be
Z(p_1,p_2,p_3) = B^{-1/2}(p_1)A^{-1/2}(p_2) A^{-1/2}(p_3)
\label{zed}
\ee
This definition proves very useful in 
reducing the 
complexity of the set of equations (\ref{SD1})-(\ref{SD3}), 
under certain approximations to be discussed 
in detail in the next sections. In addition, 
the quantity
\be 
g_R(p_1,p_2,p_3) \equiv e {\hat G}(p_1,p_2,p_3)
\label{runcoupl}
\ee
provides a 
natural generalization of the concept of the
running 
or effective charge in the
context of superrenormalizable gauge
theories \cite{jmcres}, such as $QED_3$.
This running of the coupling 
should be understood as a 
Wilsonian rather than Gell-Mann-Low type,  
in the sense that 
it is not associated with ultraviolet infinities; 
instead, it expresses a  non-trivial infrared structure of the 
theory~\cite{am}. 

Note that in four-dimensional $QED$ 
($QED_4$) the effective charge $e^2_{\rm eff}$ is 
defined in terms of the photon vacuum polarisation as:
\be
e^2_{\rm eff}(q^{2}) = e^2 [1 + \Pi(q^{2})]^{-1}, 
\label{pertphot}
\ee
and 
is a gauge-, scale-, and scheme-independent quantity \cite{JW}
to all orders in perturbation theory. $e^2_{\rm eff}$ depends explicitly
on $q^2$ {\it and} the masses of the
fermions inside the vacuum polarisation loop.
 In the limit where
the fermion masses can be neglected, 
$e^2_{\rm eff}(q^{2})$
coincides with the running coupling obtained by the
$\beta$ function of $QED_4$, i.e. the solution of the usual 
renormalization-group differential equation.
An advantage of the definition given in (\ref{runcoupl}) 
is that it captures the running coupling even in the
case of scalar theories \cite{cm}, 
where, due to the absence of Ward-Takahashi identities, 
the role of the gauge boson self-energies 
is not as prominent as in 
gauge theories. As we shall see in section 5, the interpretation
of $g_R$ defined in (\ref{runcoupl}) as a running coupling
is also justified by the form of the SD for the fermion mass gap.

The equation for the semi-amputated vertex $\hat{G}$
may be obtained from (\ref{SD3}) by multiplying both sides by the
factor $Z(p_1,p_2,p_3)$, i.e.
\be  
{\hat G}(p_1,p_2,p_3)\gamma_{\mu} = 
Z(p_1,p_2,p_3)\gamma_{\mu} 
-e^2  \int \frac{d^3k}{(2\pi)^3} {\hat G}^3 
\gamma^{\alpha}\frac{1}{\nd{k}-\nd{p_1}}\gamma_{\mu} \frac{1}{\nd{k}}
\gamma_{\alpha}\frac{1}{(k+p_2)^2} 
\label{SDamp}
\ee
where
\be
{\hat G}^3 \equiv {\hat G}(p_3,k+p_2,p_1-k){\hat G}(p_1,-k,k-p_1)
{\hat G}(p_2,k,-k-p_2)
\ee
In what follows we shall restrict ourselves to the case where the photon 
momentum is vanishingly small, and thus one is left with a 
single momentum scale $p$. 
One can then define a 
renormalization-group $\beta$ function from this ``running'' coupling 
$G(p)$ by setting
\be
   \beta \equiv p \frac{d}{d p}{\hat G}(p) 
\label{beta}
\ee
In order to further simplify the SD equation for $G(p)$  
we make the additional 
approximation that ${\hat G}^3 = {\hat G}^3(k)$, i.e. a cubic power 
of a single ${\hat G}(k)$ 
depending
only on  the integration variable $k$. This approximation will be 
justified by the self-consistency of the solutions. 
  
Carrying out the 
gamma-matrix algebra using the formulae
$ \gamma_{\mu}\gamma_{\mu} = -d $, and
$\gamma_{\mu}\gamma_{\rho}\gamma_{\mu} = (d-2)\gamma_{\rho}$
valid for $4\times 4$ gamma matrices in $d(=3)$-dimensional Euclidean space,
one obtains in a straightforward manner:
\be
{\hat G}(p) = 
Z(p) + \frac{1}{3}e^2 \int \frac{d^3k}{(2\pi)^3} {\hat G}^3 (k) 
\frac{1}{k^2}\frac{1}{(k-p)^2}
\label{asymptotic}
\ee
Several remarks are now in order. First,
one observes that $Z(p) \rightarrow 1$ for 
$p \rightarrow \infty$, where perturbation theory is valid.  
This is inferred 
from the fact that in such a case, as can be readily verified,  
the functions  $A(p), B(p) \rightarrow 1 
+ {\cal O}(e^2/p)$. 
In addition, in the ultraviolet region, $p \rightarrow \infty$, 
gauge invariance requires ${\cal G}(p) \sim A(p)$. 
Second, from (\ref{asymptotic}) one observes that,
if ${\hat G}$ stays positive, which is expected for any physical 
theory, then, 
as a result of the positivity of 
the integrand, ${\hat G(p)} \ge Z(p)$ for any $p$. 
Thus, one has the following basic properties of ${\hat G}(p)$,
which stem directly from the integral equation (\ref{asymptotic}):
\be
  {\hat G}(p) \ge Z(p),~ {\rm for~all}~ p, \quad ; \quad {\hat G}(p) 
\sim B^{-1/2}(p)
\rightarrow 1, \quad p \rightarrow \infty
\label{ascond}
\ee
Notice that ${\hat G}(p)$ in the ultraviolet is thereby given 
by $\left(1 + \Pi (p)\right)^{-1/2}$ as in $QED_4$. 
This is the perturbative result, which,  
as we shall see later, is modified 
non-trivially in the infrared,  
in a way consistent with gauge invariance. 
Moreover, as we shall see later, 
the
self-consistency of the approximations employed 
will require 
${\hat G}(0) >> \sqrt{3/2}$.

Our next assumption is that in
the infrared regime, 
$k/\alpha << 1$, 
which is of interest to us here,  
the effects of the inhomogeneous term 
$Z(p)$
can be ignored.
This assumption will be justified later on,
when we consider a non-trivial self-consistency check 
of the solutions. We now remark that by 
ignoring the effects of $Z(p)$ one can decouple
the equation for the (gauge-invariant) {\it amputated} vertex
from the equations for $A(p)$,$B(p)$. 
As we shall discuss in subsequent sections, these latter 
equations also decouple from each other, depending only 
on the vertex function ${\hat G}(p)$.

Because of this, we commence our analysis 
from the  SD equation for the vertex ${\hat G}(p)$,
which we solve upon ignoring the effects of the inhomogeneous $Z(p)$ term. 
Thus we arrive at the homogeneous equation
\be
     {\hat G}(p) = 
\frac{1}{3}e^2 \int \frac{d^3k}{(2\pi)^3} {\hat G}^3 (k) 
\frac{1}{k^2}\frac{1}{(k-p)^2}
\label{ampapp}
\ee
This integral equation involves only one unknown function, namely ${\hat G}$,
which must be self-consistently determined.
Note that this equation is invariant under the rescaling 
${\hat G} \rightarrow {\hat G}/e$. This indicates 
a straightforward extension of the analysis to a large-$N$ treatment,
given that $N$ can be absorbed in a redefinition of $e^2$.

It is easy to see that, written 
in the form (\ref{ampapp}),
the equation does not admit {\it physically acceptable} solutions, i.e. 
solutions with ${\hat G} \ge 0$ and {\it finite}~\footnote{Solutions 
that blow up in any point of the integration region are 
discarded.}. Indeed, setting  
$p =0$ one obtains after the (trivial) angular integration~\footnote{These
arguments remain unaffected even in the presence of an 
inhomogeneous term $Z(p)$, such that $Z(0)$ is non-negative 
and finite. 
The non-negative nature of $Z(p)$ (\ref{zed}) stems 
from that of $A(p)$, which  
is guaranteed from general renormalization-group arguments~\cite{books}.}: 
\be
       {\hat G}(0) = \frac{e^2}{12\pi^2} \int _0^\infty \frac{dk}{k^2} 
{\hat G}^3(k) 
\label{limito}
\ee
Finiteness of ${\hat G}(0)$ requires that the integrand of the right hand side 
of (\ref{limito}) converges at $y \rightarrow 0~{\rm and}~ \infty$.
The ultaviolet limit does not present a problem, because the 
kernel vanishes like $y^{-2}$, which is consistent with the 
superrenormalizability of the theory as well as the fact that 
the amputated vertex tends to 1. 
In the infrared limit $y \rightarrow 0$, however, 
the kernel blows up. 
For the integral to remain finite at that point, as required
by the finiteness assumption for ${\hat G}(0)$, 
$G^3(y)$ must approach zero as $y^\alpha,~\alpha > 1/3$, thereby 
implying that ${\hat G}(0) =0$. 
However for that to happen the integrand in (\ref{limito}) 
must change sign, which would in turn imply that ${\hat G}(y)$ itself  
must change sign somewhere in $y$.  
According to our assumption above this 
is not a physically acceptable situation.

To show rigorously that there are no 
{\it physically} acceptable solutions 
leading to a 
non-trivial infrared structure
we next 
convert 
the integral equation into a non-linear differential equation 
of the type known in the mathematical literature as
{\it Emden-Fowler} equation~\cite{emden,kamke}.
To this end, we perform the angular integration in (\ref{ampapp}), 
to arrive at the equation:
\be 
  {\hat G}(p) = \frac{2}{3\pi^2} \frac{\alpha}{p} \int _0^\infty \frac{dk}{k} 
{\hat G}(k)^3 {\rm ln}|\frac{k+p}{k-p}|
\label{v2}
\ee
where we have set $e^2 \equiv 8\alpha$ to make contact with the 
usual large-$N$ definition~\cite{app}. For us, however, 
the number of fermion flavours is not assumed to be  necessarily 
large. In fact, for brevity 
we set $N=1$ (in a four-component notation
for the fermions) throughout this work. 
Next we introduce the dimensionless variables 
$x \equiv p/\alpha$ and $y\equiv k/\alpha$. Since we
are interested in the infrared behaviour of the model we consider
the limit  
$x <<1 $, for which one obtains by expanding the logarithms
in the integrand: 
\bea 
&~&   {\hat G}(x) =\frac{2}{3\pi^2 x } \int _0^\infty 
\frac{dy}{y} {\hat G}^3 (y) 
{\rm ln}|\frac{y+x}{y-x}| \simeq \nn \\
&~& \frac{4}{3\pi^2 x^2}\int _0^x dy {\hat G}(y)^3 + \frac{4}{3\pi^2 }
\int _x^\infty \frac{dy}{y^2} {\hat G}^3 (y) 
\label{v4}
\eea
Differentiating appropriately with respect to 
$x$, we arrive at the following differential equation for small $x$: 
\be 
x^3 \frac{d^2 {\hat G}}{d x^2} + 3 x^2 \frac{d {\hat G}}{d x} + 
\frac{8}{3\pi^2}{\hat G}^3(x) =0  
\qquad x << 1 
\label{fcaleq}
\ee

It is convenient to rescale ${\hat G}$ by setting:
\be
    G \equiv \sqrt{\frac{8}{3\pi^2}} {\hat G} 
\label{definighat}
\ee
Then, eq. (\ref{fcaleq}) becomes:
\be
x^3 \frac{d^2 G}{d x^2} + 3 x^2 \frac{d G}{d x} + 
G^3(x) =0 , \qquad x << 1
\label{fcaleq2}
\ee
Upon the change of variables
$\xi = \frac{1}{2 x^2}$, $G=2^{3/4}~\eta(\xi)$, 
the equation becomes of Emden-Fowler type~\cite{emden,kamke}:
\be
    \frac{d^2}{d \xi^2} \eta (\xi) + \xi^{-3/2} \eta^3 (\xi) =0, \qquad 
\xi \rightarrow +\infty 
\label{emden2}
\ee    
As discussed in the mathematical literature~\cite{emden},
the {\it only non-trivial 
real solutions}  of (\ref{emden2}), as $\xi \rightarrow +\infty$,
are {\it oscillatory}
about zero of simple sinusoidal form, which however 
oscillate infinitely  rapidly as $x \rightarrow 0$.
The amplitude of the above solutions behaves like $x^{-1/2}$, for $x << 1$. 
We interprete 
this behaviour as indicating 
an {\it instability} of the massless-fermion ground state.

It is interesting to notice that the only 
power-law solution of (\ref{fcaleq2}) 
for $x << 1$, is {\it purely imaginary}, i.e.  
\be
      G(x) = i\frac{\sqrt{5}}{2}x^{1/2}, ~\qquad x << 1
\label{fp}
\ee
This 
solution would imply a `trivial infrared fixed point structure'
given that its associated $\beta$ function vanishes at $x=0$. 
However, 
the fact that (\ref{fp}) is {\it purely imaginary }
would again suggest instability.

The above analysis 
constitutes a rigorous proof that, within the context of {\it proper}
(i.e. finite and with finite-derivatives) solutions, and modulo the 
approximations discussed, 
no non-trivial infrared-fixed point is possible in $QED_3$ in the absence 
of an infrared  cut-off. This was conjectured  in ref. \cite{am}, 
but here we have given an analytic proof.  
This motivates one to look for the existence
of a possible non-trivial infrared fixed-point structure 
in the presence of fermion and/or photon masses. 
In the next section we shall discuss the case 
when the fermions develop a mass $m$. As we shall show,
the existence of a non-trivial infrared fixed point 
is guaranteed due to the form of the resulting equations.

\setcounter{equation}{0}

\section{Equation for the Vertex in the case of non-zero fermion mass}

As a first kind of infrared cut-off in the integral equation 
(\ref{ampapp}) we shall consider the case of a fermion 
mass gap $m(p)=\frac{\Sigma (p)}{A(p)}$, where $\Sigma (p)$ is 
the fermion self energy. 
In that case the fermion propagator $S_F$ becomes:
\be
     S_F (k) = \frac{i}{A(k) \left( \nd{k} + m_f(k)\right)}
\label{fermass}
\ee
For our purposes below we assume that $m_f(p) \simeq m_f(0) \equiv m_f \ne 0$.
In that case the integral equation (\ref{ampapp}) becomes:
\be
   {\hat G}(p) =   
\frac{1}{3}e^2 \int \frac{d^3k}{(2\pi)^3} {\hat G}^3 (k) 
\frac{1}{(k^2+ m_f^2 )(k-p)^2} + \frac{2m_f^2}{3}
\int \frac{d^3k}{(2\pi)^3}{\hat G}^3 (k)
\frac{1}{(k^2 + m_f^2)^2 (k-p)^2}
\label{ineq2}
\ee
Notice that the effects of the fermion mass are not given simply
by just adding a mass squared term in the fermion denominators, 
but they result in additional structures in the integral equation. 

Performing the angular integrations 
one arrives at: 
\be
{\hat G}(x) = \frac{2}{3\pi^2 x}\int dy f(y) 
{\rm ln}\left|\frac{y+x}{y-x}\right|{\hat G}^3(y) 
\label{equntitl}
\ee
where $x \equiv p/\alpha$, $m \equiv m_f/\alpha$ are dimensionless,
and 
\be
  f(y) \equiv y \frac{y^2 + 3m^2}{(y^2 + m^2)^2} \ge 0 
\label{fdef}
\ee
Differentiation with respect to $x$ yields:
\be
 x\frac{d}{d x}{\hat G}(x) =-\frac{2}{3\pi^2 x}\int _0^\infty 
dy f(y) \left( {\rm ln}\left|\frac{y+x}{y-x}\right|+\frac{2xy}{x^2-y^2}\right)
{\hat G}^3(y) 
\label{kernel}
\ee
One observes that formally as $x\rightarrow 0$ the right-hand-side vanishes,
provided that ${\hat G}$ is finite. This indicates the existence 
of a fixed point. As we shall show below this is confirmed
analytically by converting the integral equation into 
a non-linear differential equation. 

An additional feature which one would have hoped 
to study already at 
the level of the integral equation (\ref{kernel}) is the monotonicity 
of ${\hat G}$. 
Unfortunately the kernel in (\ref{kernel}) is not manifestly 
positive to allow for such an analytic proof at the level of the 
integral equation for generic values of $x$, and one has to 
resort to numerical treatments, which fall beyond the scope
of this article. 
However one can already infer from (\ref{kernel}) 
that,  
for high momenta $x >> 1$, 
a monotonically decreasing ${\hat G}$ 
is consistent with the expectation that in this regime 
${\hat G}$ 
is essentially given by its perturbative expression which 
asymptotes to $1$. The analysis is omitted because 
it is straightforward. 

For low momenta, on the other hand, 
the behaviour of ${\hat G}(x)$ will also be shown to 
be monotonically decreasing, 
starting from a non trivial fixed point.
This will be achieved by converting the integral equation 
into a differential one.  
Unfortunately, at present, we cannot analytically derive 
the monotonicity for intermediate momenta. 

To derive the differential equation from (\ref{equntitl}) 
we follow a similar analysis to the one leading to (\ref{v4}). 
First, one expands the logarithms 
for small $x << 1$, thereby writing the equation as: 
\bea
&~& {\hat G}(x) \simeq 
\frac{4}{3\pi^2 x^2}\int _0^x dy {\hat G}(y)^3 \frac{y^2}{y^2 + m^2} + 
\frac{4}{3\pi^2 }
\int _x^\infty \frac{dy}{y^2 + m^2} {\hat G}^3 (y) + \nn \\
&~&  
\frac{8m^2}{3\pi^2 x^2}\int _0^x dy {\hat G}(y)^3 \frac{y^2}{(y^2 + m^2)^2} + 
\frac{8m^2}{3\pi^2 }
\int _x^\infty \frac{dy}{(y^2 + m^2)^2} {\hat G}^3 (y) 
\label{eq2}
\eea
Differentiating with respect to $x$ one arrives at:
\be
x(x^2 + m^2)^2 \frac{d^2}{d x^2}{\hat G}(x) + 
3(x^2 + m^2)^2\frac{d}{dx}{\hat G}(x) + 
\frac{8}{3\pi^2}(x^2 + 3m^2){\hat G}^3(x) =0
\label{emdenmass}
\ee
The above equation can be solved numerically, 
to which we will come later on. 
However,  
in the infrared region 
$x << m$ the equation 
accepts an analytic treatment, as we discuss below. 
In this region the equation (\ref{emdenmass}) is approximated by:
\be
    x \frac{d^2}{dx^2}{\hat G}(x) + 3 \frac{d}{d x}{\hat G}(x) + 
\frac{8}{\pi^2m^2}{\hat G}^3(x)=0
\label{papmav}
\ee
It is immediate to see that 
a special power-law solution is given by (for positive $G(x)$)
by:
\be
 {\hat G}(x) =m \pi \frac{\sqrt{3}}{4\sqrt{2}}x^{-1/2}
\label{ss}
\ee
Notice the infrared divergence of this type of solutions {\it even}
in the presence of a (bare) fermion mass. 
The associated renormalization-group $\beta$ function (\ref{beta})
for this case reads:
\be
     \beta (x) = -\frac{1}{2}{\hat G} \sim x^{-1/2} 
\rightarrow +\infty,~{\rm as}~x \rightarrow 0
\label{beta2}
\ee
indicating the absence of an infrared fixed point. 
The associated operator appears to be {\it relevant} 
(negative scaling dimension), which implies the possibility that the theory 
be driven to a non-trivial fixed point. 

However, in the infrared regime $x << 1$, 
one can find a different type of solution~\cite{emden}:
\be
    {\hat  G} =  m \pi \frac{\sqrt{3}}{2\sqrt{2}}\frac{c}{1 + c^2 x}, \qquad x \rightarrow 0
\label{es}
\ee
where $c$ is a constant of integration to be fixed 
by the boundary condition at $x=0$ implied by the integral equation,
to be discussed later on. For physical solutions $c$ is assumed positive.

This type of solutions has a renormalization-group $\beta$-function
(\ref{beta}) of the form:
\be
     \beta = -{\hat G}(x) + \frac{2\sqrt{2}}{\sqrt{3}\pi m c}{\hat G}^2(x) \sim 
-\frac{x}{\left(1 + c^2 x\right)^2} \rightarrow 0, \qquad x \rightarrow 0
\label{beta3}
\ee
from which we observe the existence of a non-trivial (non-perturbative) 
infrared fixed 
point at ${\hat G}^* = \frac{\pi m \sqrt{3}c}{2\sqrt{2}} >0$.  
Such a fixed point 
is the result of the dynamical generation of a 
parity-invariant, chiral-symmetry breaking 
fermion mass~\cite{app}, indicating 
the connection of the phenomenon of chiral symmetry breaking
in $QED_3$ to a non-trivial infrared fixed point structure,
in agreement with the expectations of ref. \cite{am}.

The non-trivial fixed-point solution (\ref{es}) 
{\it is not} compatible with the integral equation (\ref{ineq2}) 
for {\it any} value of the fermion mass $m$. 
Indeed one can derive a 
{\it boundary condition} for ${\hat G}(0)$ from (\ref{ineq2}), which reads:
\be
 {\hat G}(0) 
= \frac{4}{3\pi^2}\int _0^\infty dy \frac{1}{y^2 + m^2} {\hat G}^3(y) 
+ \frac{8m^2}{3\pi^2}\int _0^\infty dy \frac{1}{(y^2 + m^2)^2}{\hat G}^3(y)
\label{bc2}
\ee
In contrast to the massless case (\ref{limito}), ${\hat G}(0)$ 
is now a finite constant, $\pi m c \sqrt{3/8}$, as seen from (\ref{es}), 
and this allows for a compatibility of the solution (\ref{es}) 
with (\ref{bc2}), {\it provided} that $m {\hat G}(0)$ 
satisfies certain conditions to be specified below.

To this end,  
we split the $y$-integration in (\ref{bc2}) into
two intervals: (i) $y \in [0, m)$, where the form (\ref{es}) 
is valid to a good approximation, and (ii) $y \in [m, \infty)$, 
where we approximate ${\hat G}(x)$ by its perturbative asymptotic
value ${\hat G} \simeq 1$. By this latter approximation we 
overestimate the actual value of the integration, 
given that ${\hat G}$ is actually 
slightly smaller than unity for finite $p$, approaching it 
only asymptotically (c.f. (\ref{ascond})). 
However, this is sufficient 
for our qualitative purposes of demonstrating the 
existence of 
constraints 
on the fermion mass implied by the boundary condition (\ref{bc2}).    
 
\begin{figure}[htb]
\epsfxsize=3in
\bigskip
\centerline{\epsffile{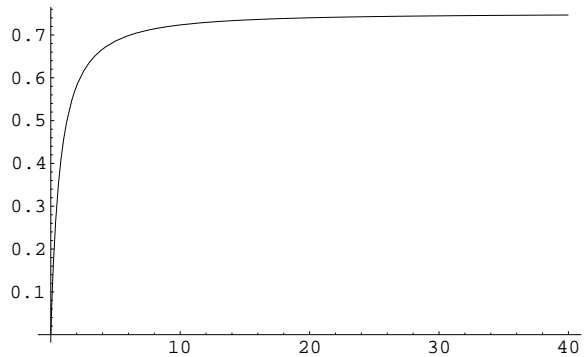}}
\caption{\it\baselineskip=12pt 
Plot of the function $f(z)=\frac{z}{2}\int _0^1 dy \frac{1}{y^2 + 1} \frac{1}{(1 + 
z y)^3}
\left( 1 + \frac{2}{y^2 + 1}\right)$, where $z=mc^2$. The function 
asymptotes rapidly to $3/4$.}
\bigskip
\label{fig2}\end{figure}

With these in mind, the boundary condition (\ref{bc2}) reduces to:
\be
1 = \frac{mc^2}{2}\int _0^1 dy \frac{1}{y^2 + 1} \frac{1}{(1 + mc^2 y)^3}
\left( 1 + \frac{2}{y^2 + 1}\right) 
+ \frac{4}{3\pi^2 m^2c}\left(1 - \frac{1}{\pi}\right)\sqrt{\frac{2}{3}}
\label{gap}
\ee
To obtain the condition imposed on $m$ by the 
boundary condition (\ref{bc2}) it suffices to 
observe that 
the first term 
on the right-hand-side
is a function of $mc^2$ alone, and that,  
after the (elementary) $y$ integration, the 
resulting function of $mc^2$ asymptotes 
rapidly to the value $3/4$ 
(see figure \ref{fig2}).

This implies the following inequality: 
\be
   m {\hat G}(0) < \frac{8}{3\pi}\left(1 - \frac{1}{\pi}\right)
\label{cond}
\ee
As already mentioned this bound is overestimated, given that 
in the actual situation the function ${\hat G}(y)$ is not exactly $1$
immediately after the region $y \ge m$.

We next remark that $m$ should actually be determined 
self-consistently from a solution of the pertinent gap equation.
This will be done in section 5. 
However, at the moment, and for completeness,   
we shall assume that $m$ is determined by 
its approximate form derived within the context 
of a large-$N$ treatment~\cite{app,dor}. 
Compatibility of the dynamical solution with the constraint 
(\ref{cond}) 
will then lead to further restrictions on the range of the 
allowed masses $m$. As we shall see in section 5, 
there is good agreement 
between the allowed fermion-mass ranges, 
obtained 
within the context of a large-$N$ treatment, 
and those obtained 
from a 
self-consistent solution of the mass gap 
equation within our approach. 

\begin{figure}[htb]
\epsfxsize=3in
\bigskip
\centerline{\epsffile{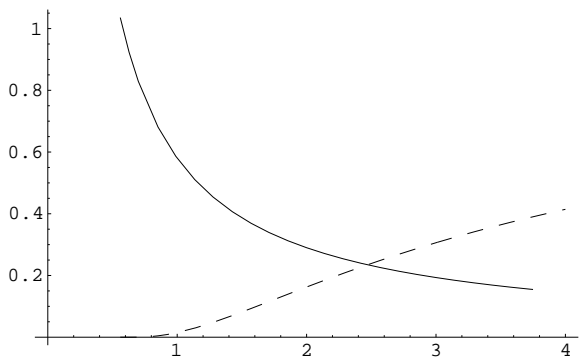}}
\caption{\it\baselineskip=12pt 
Fermion mass versus the infrared-value 
of the coupling ${\hat G}(0)$. The solid curve 
represents the condition 
derived from the integral equation for the vertex, 
whereas the dashed 
line represents the solution obtained from 
the standard gap equation in the 
large-N treatment.}
\bigskip
\label{fig3}\end{figure}

In the context of a large $N$ treatment, and 
to leading order in $1/N$ resummation, the following solution
for the dynamically-generated $m$  
is found~\cite{app,dor}, 
\be
    m \sim {\cal O}(1) {\rm exp}\left(-\frac{2\pi}{\sqrt{\frac{g^2}{g_c^2}-1}}\right)
\label{appelquist}
\ee
where $g_c^2 =\frac{\pi^2}{32}$ is the critical coupling, above which
dynamical mass generation occurs~\cite{app}. 
Compatibility of the solution (\ref{appelquist}) 
with the constraint
(\ref{cond})
implies the existence of an {\it upper bound} on fermion masses,  
$m < m_{max}$, where $m_{max}$ is defined
through the intersection of the curves (\ref{cond}) and (\ref{appelquist})
in the $\left(m,g\right)$ plane (see fig. \ref{fig3}). 
This yields
$m_{max} \simeq 0.3$.

On the other hand, for large momenta, we know that ${\hat G} \rightarrow 1$.
Physically one expects
{\it a monotonic decrease}  of ${\hat G}(x)$ over the {\it entire}
range of $x \in [0, \infty)$. This would occur in our case 
if and only if ${\hat G}(0) > 1$, which,
in the context of the large-$N$ result of \cite{app,dor}, implies
a {\it minimum} bound for the fermion masses   
$m > m_{min}\simeq 0.03$. Actually, as we shall argue 
in the next section, ${\hat G}(0)$ should be comfortably 
larger than $\sqrt{3/2}$ 
for self-consistency of our approximations. 

\begin{figure}[htb]
\epsfxsize=3in
\bigskip
\centerline{\epsffile{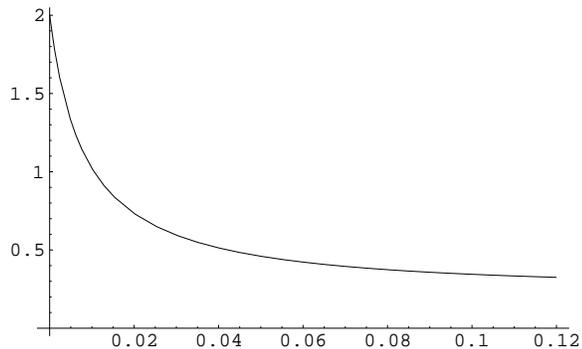}}
\caption{\it\baselineskip=12pt 
Numerical solution of the equation (\ref{emdenmass}) versus $p/\alpha$,
for a typical set of values $m \sim 0.1$ and ${\hat G}(0)=2$;
the solution decreases 
monotonically and asymptotes quickly to a positive constant value.}
\bigskip
\label{fig4}\end{figure}

Hence, we see that the monotonicity of the
running coupling can be achieved 
in the context of a large-$N$ treatment, 
if 
the mass $m$ lies in the following regime:
\be 
      0.03 \lsim m \lsim 0.3 
\label{zone}
\ee
or equivalently if the coupling at the infrared 
point ${\hat G}(0)$ is restricted in the regime (see fig. \ref{fig2}):
\be
           1 < {\hat G}(0) < 2.5,\qquad 
\label{regionc}
\ee
At this point it is useful to turn to a numerical 
study of the equation (\ref{emdenmass}), supplemented with 
the boundary conditions imposed by 
the solutions of the form (\ref{es}),
specifically:
\be
    {\hat G}(0)=m\frac{\pi}{2}\sqrt{\frac{3}{2}}c \qquad; 
\qquad {\hat G}'(0)=-\frac{8 {\hat G}^3(0)}{3\pi^2 m^2}
\label{bc3}
\ee
where the constant $c$ should be chosen in such a way that the bound 
(\ref{cond}) be satisfied.
The numerical solution of (\ref{emdenmass}) is given in 
figure \ref{fig4}, for a typical case, where $m=0.1$, and ${\hat G}(0)=2$. 
As we observe, the figure clearly demonstrates 
{\it the monotonic decrease} of ${\hat G}$ in the small $x$ interval, where
equation (\ref{emdenmass}) is valid. 
We also note that the solution asymptotes quickly 
to a constant positive value,
smaller than $1$ in contrast to (\ref{ascond}), which would require
${\hat G}(x) \sim 1$, for $x >> 1$.  
This, however, presents 
no contradiction, because 
equation (\ref{emdenmass}) 
was based on ignoring the inhomogeneous term $Z(x)$.
As one goes to a region of larger $x$, 
this assumption is no longer valid.  
In that case  
the inhomogeneous term $Z(x)$ becomes important, 
and  
equation (\ref{emdenmass})   
receives additive (positive) 
contributions, resulting in ${\hat G}(p) \sim 1$, for large $p$, as 
required by 
(\ref{ascond}).

\setcounter{equation}{0}
\section{Equations for Photon and Fermion self-energies}

As a consistency check of our assumption about ignoring 
the effects 
of the inhomogeneous term $Z(p)$ as $p \rightarrow 0$,
next to  
${\hat G}$,
we now turn our attention to the 
equations that determine the fermion and photon self-energies 
$A(p)$
and $B(p)$ respectively.
As shown below, upon neglecting $Z(p)$ in the infrared,
these equations decouple 
from each other and both functions 
can be determined solely from knowledge of ${\hat G}(x)$. 
It should also be stressed that in the massless case the
system does not admit a self-consistent solution. 
In contrast, 
the presence a fermion mass term 
changes the situation drastically by yielding self-consistent 
solutions for $A(p),B(p)$ which are such that 
$Z(p) \rightarrow 3\sqrt{6}/5$, but 
at a rate slower than the one with which  
${\hat G}$ approaches ${\hat G}(0)$ as $x \rightarrow 0$. 
Thus the approximation of 
ignoring the effects of $Z(x)$ in the region $x << 1$ 
is qualitatively correct. 
This is a highly non-trivial check 
of our approach, and 
justifies fully the approximations used above.  

To this end,  
we begin 
from the integral equation for
$A(x)$, 
which reads: 
\be
A(p)~\nd{p} = \nd{p} - e^2\int \frac{d^3k}{(2\pi)^3}{\cal G}^2(k)
\gamma _\mu \frac{i}{A(k)\nd{k}}\gamma _\nu \frac{\delta _{\mu\nu}}{B(k-p)(k-p)^2}
\label{fermiself}
\ee
which in terms of the semi-amputated vertex ${\hat G}$ becomes:
\be
A(p)~\nd{p} = \nd{p} - e^2A(p)\int \frac{d^3k}{(2\pi)^3}{\hat G}^2(k)
\frac{\nd{k}}{k^2(k-p)^2}
\label{fermiself2}
\ee
To arrive at the above equation we have carried out the $\gamma$-matrix 
algebra and we have used the approximation that 
${\hat G}(k,p) ={\hat G}(k)$. This approximation is equivalent to the one made 
for ${\hat G}$ in (\ref{SDamp}). It will be justified below, 
by the self-consistency of the solution. 

In terms of the dimensionless $x,m$ parameters 
introduced previously, the equation (\ref{fermiself2}) 
becomes:
\be
x^4=A^{-1}(x)x^4  -\frac{2}{\pi^2}\int _0^x dy \frac{y^4 {\hat G}^2 (y)}{y^2 + m^2} - \frac{2 x^4}{\pi^2}\int _x^1 dy \frac{{\hat G}^2 (y)}{y^2 + m^2}
\label{eqwf}
\ee
Differentiating twice with respect to $x$ 
one obtains the following differential equation: 
\be
  \frac{d}{dx}\left[x^{-3} \frac{d}{dx}\left(x^4 A^{-1}(x)\right)\right] 
+ \frac{8}{\pi^2}\frac{{\hat G}^2 (x)}{x^2 + m^2} = 0
\label{eqdiffwf}
\ee
Restricting our attention to the infrared region $x \rightarrow 0$, one may 
ignore $x^2 $ in front of $m^2$, and use the form 
(\ref{es}) 
for ${\hat G}(x)$. After elementary integrations one then finds
for $A(x)^{-1}$:
\bea
&~&A^{-1}(x) =\frac{3}{2}\left[-\frac{2}{c^7}x^{-3} + \frac{1}{c^5}x^{-2} - \frac{2}{3c^3}x^{-1}
+ \frac{2}{c^9}x^{-4}{\rm ln}\left(1 + c^2 x \right)\right] + \frac{1}{4}c' \simeq 
\nn \\
&~&\frac{1}{4}(c'-\frac{3}{c}) + \frac{3}{5}cx -\frac{1}{2}c^3x^2
+  {\cal O}(x^5), \qquad x \rightarrow 0
\label{awf}
\eea
where $c>0$ is the integration constant of the solution (\ref{es}),
which obeys the restriction (\ref{cond}). We choose the constant 
of integration $c'=3/c$, so that
\be
      A(x)^{-1} \simeq \frac{3}{5}c x - \frac{1}{2}c^3 x^2 + \dots 
\label{awff}
\ee 
as $x \rightarrow 0$. As  we shall show later, this choice 
is compatible with the boundary behaviour $Z(x) \rightarrow {\rm const}$,
as $x \rightarrow 0$, 
advocated for the inhomogeneous term.

Next we turn our attention to the equation for the photon function $B(p)$,
in the presence of a fermion mass. 
Following similar steps to the ones leading to (\ref{fermiself2}),  
the pertinent integral equation reads: 
\be
B(p)\left(p^2 \delta _{\mu\nu} - p_\mu p_\nu \right)= 
p^2 \delta _{\mu\nu} - p_\mu p_\nu 
+B(p)e^2 \int \frac{d^3k}{(2\pi)^3} {\hat G}^2(k){\rm Tr}\left[\gamma _\mu \frac{\nd{k}}{k^2}
\gamma _\nu \frac{\nd{k}-\nd{p}}{(k-p)^2}\right]  
\label{photonself2}
\ee
To trasnform the above equation into a scalar 
equation for $B(p)$ it is necessary to take the trace on both sides.  
In doing so one assumes that the integral term on the 
right-hand-side is also transverse. This is expected from gauge 
invariance. In our truncated scheme it can be demonstrated  
that this is indeed the case, provided one makes our working hypothesis
that ${\hat G}(k,p)={\hat G}(k)$. Specifically if one contracts 
both sides of (\ref{photonself2}) by $p^\mu$, the right-hand-side
vanishes (Ward identity for the photon) 
only upon making the above assumption for the semi-amputated
vertex, and shifting appropriately the integration variable. 
 
After standard manipulations, similar to the ones above, 
one arrives at: 
\be 
   \frac{d}{dx}\left[x^{-1}\frac{d}{dx}\left(x^{-1}\frac{d}{dx}\left(x^4 B^{-1}\right)\right)\right] + \frac{32}{\pi^2}\frac{1}{x^2 + m^2}{\hat G}^2 (x) = 0, \qquad x \rightarrow 0
\label{photonwf}
\ee
Making the same approximations in the infrared $x << 1$ as in the case
of $A(x)$, and substituting the Emden's solution (\ref{es}) 
for ${\hat G}$ on the right-hand-side of (\ref{photonwf}), we obtain
after some elementary integrations, upon setting their 
constants but one ($c_1$) to zero:
\be
   x^4 B^{-1}(x)=\frac{6}{c^2}x^2 -\frac{6}{c^6}x + \frac{3}{c^4}x^2
+ \frac{6}{c^8}{\rm ln}\left(1 + c^2 x\right) - \frac{6x^2}{c^4}
{\rm ln}\left(1 + c^2 x\right) + \frac{1}{8}c_1 x^4 
\label{betacapital}
\ee
which for $x << 1$ yields:
\be
   B^{-1}(x) \simeq \frac{6}{c^2}x^{-2} \left[1 - \frac{2}{3} x + 
 \left(\frac{c_1}{8}+ \frac{3}{2}\right)\frac{c^2}{6} x^2 + \dots \right]
\label{photonwff}
\ee
This form of $B(x)$ gives rise to 
the following form for the 
photon propagator in the infrared $k \rightarrow 0$,
\be
   \Delta _{\mu\nu} \sim \frac{\delta _{\mu\nu}}{k^4} , \qquad k \rightarrow 0
\label{photinfr}
\ee
Notice, as a consistency check, that the photon continues to be massless
in the chiral-symmetry broken phase. 
The corresponding static effective potential 
is given by the appropriate 
Fourier transform of the $00$ (temporal) 
component of the photon propagator for $k_0=0$. In the case 
(\ref{photinfr}) this yields formally an effective 
potential scaling like $R^2$ for large distances $R$,
suggestive  
of confining behaviour~\footnote{However, 
this is 
a formal result, given that 
the corresponding momentum integrals are infrared divergent 
and hence need proper regularization, which falls beyond our scope here. 
For standard treatments see 
discussions in \cite{conf}.}.
The fact that this behaviour is found 
in the chiral-symmetry broken case is consistent 
with general (four-dimensional) arguments~\cite{conf,pc} that 
confinement is a sufficient but not necessary condition 
for chiral-symmetry breaking.

An important feature of the expressions (\ref{awff}) and (\ref{photonwf}) 
is that, although they are derived only in case where 
a fermion mass $m \ne 0$, however they do not explicitly depend on
the magnitude of the mass. 
From (\ref{awff}) and (\ref{photonwff}) one obtains for 
$Z(x)$ in the infrared region $x \rightarrow 0$:
\be
Z(x) =B^{-1/2}(x)A^{-1}(x) \simeq 
\frac{3\sqrt{6}}{5}-\frac{2\sqrt{6}}{5}x \left( 1 + \frac{5}{4}c^2\right)
+ {\cal O}(x^3)\dots, \qquad x \rightarrow 0  
\label{vanishingZ}
\ee
where we have 
chosen the constant of 
integration $c_1$ such that there are no ${\cal O}(x^2)$ terms.   

In order to check under which conditions the omission of 
the $x$-dependent parts of $Z$ next to those of ${\hat G}(x)$ 
is justified, 
we must compare the corresponding linear terms in $x$ of 
both quantities. 
Using that 
\be
   {\hat G}(x) \sim \sqrt{\frac{3}{2}}\left(m\frac{\pi}{2}c -   m\frac{\pi}{2}c^3x + \dots\right), \qquad x \rightarrow 0
\label{eszero}
\ee
this requires: 
\be
\sqrt{8/3} {\hat G}^3(0) - 2 {\hat G}^2 (0) - (3\pi^2/5)m^2 >> 0 
\label{cond2}
\ee
This inequality furnishes a condition on ${\hat G}(0)$
under the  
assumption $m << 1$, which,
as we shall 
see in section 5,  
turns out to be correct. The condition is 
\be
        {\hat G}(0) >>  \sqrt{\frac{3}{2}} \simeq 1.22
\label{limit}
\ee
In the above formulas the symbol $>>$ means actually at least an order 
of magnitude. As we shall see below, such a regime for ${\hat G}(0)$ 
arises self-consistently. Note that the condition (\ref{limit}) is 
in agreement with (\ref{ascond}) given that $Z(0)=3\sqrt{6}/5 \simeq 1.47$ 
(c.f. Eq. (\ref{vanishingZ})) is only slighty larger than $\sqrt{3/2}$.  

In view of the existence of a non-zero constant value 
of $Z(x)$ in the infrared region, our analysis 
on 
the restrictions 
(\ref{cond}),(\ref{regionc}) 
has to be repeated, given that these restrictions
stem from the integral equation.
From (\ref{cond2}), and the value $Z(0)=3\sqrt{6}/5$ 
it becomes clear that the boundary condition (\ref{gap}) 
is now modified to:
\be
1 = \frac{12}{5\pi mc} + 
\frac{mc^2}{2}\int _0^1 dy \frac{1}{y^2 + 1} \frac{1}{(1 + mc^2 y)^3}
\left( 1 + \frac{2}{y^2 + 1}\right) 
+ \frac{4}{3\pi^2 m^2c}\left(1 - \frac{1}{\pi}\right)\sqrt{\frac{2}{3}}
\label{gap2}
\ee
The first term continues to asymptote to $3/4$, but now
one obtains the following restrictions on the coupling:
\be
  {\hat G}(0) < \frac{12\sqrt{6}}{5} + 
\frac{8}{3}\left(1 - \frac{1}{\pi}\right)\frac{1}{\pi m}
\label{newcond}
\ee
From the above relation it becomes clear that,  
if ${\hat G}(0) < 12\sqrt{6}/5$ there is no restriction on $m$.
However, for this regime of the couplings 
the condition (\ref{limit}), necessary  
for safely ignoring the effects of 
the inhomogeneous term $Z(p)$ in the infrared next to ${\hat G}(p)$, 
is only marginally satisfied. 
On the other hand, if one allows ${\hat G}(0)$ to exceed
this value, which is physically acceptable, 
then 
the following upper bound on $m$ as a function of ${\hat G}(0)$ is obtained:
\be
   m < \frac{\frac{8}{3\pi}\left(1 - \frac{1}{\pi}\right)}
{{\hat G}(0) - \frac{12\sqrt{6}}{5}}, \qquad {\hat G}(0) > 
\frac{12\sqrt{6}}{5} \simeq 5.88
\label{fincond}
\ee
which replaces (\ref{cond}). 

\begin{figure}[htb]
\epsfxsize=3in
\bigskip
\centerline{\epsffile{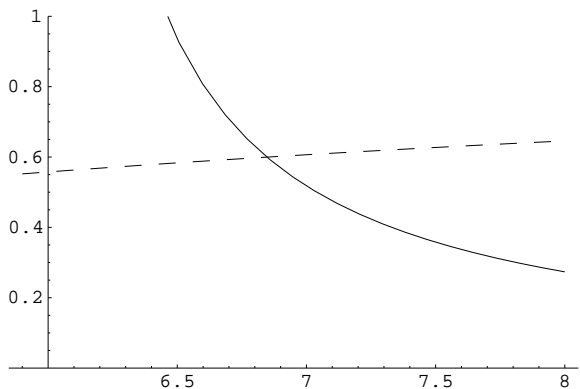}}
\caption{\it\baselineskip=12pt 
Fermion mass versus the infrared-value 
of the coupling ${\hat G}(0)$
for the case where the effects of the inhomogeneous
term (almost constant) $Z(x)$ have been taken into account in the infrared. 
The solid curve 
represents the condition 
derived from the integral equation for the vertex, 
whereas the dashed 
line represents the solution obtained from 
the standard gap equation in the 
large-N treatment.}
\bigskip
\label{app2}\end{figure}

One may repeat the comparison with the large $N$ limit results~\cite{app},
as in the previous section, 
to determine the new region of allowed values of the mass. 
The analysis is given in figure (\ref{app2}).

From fig. \ref{app2} we 
then obtain the following allowed range of the coupling 
${\hat G}(0)$:
\be
   5.88 < {\hat G}(0) < 7 
\label{regionc2}
\ee
to be compared with (\ref{regionc}) associated with the monotonicity 
requirement for ${\hat G}(p)$. 
This analysis shows that the new allowed region for the fermion masses,
in case (\ref{fincond}) is:
\be
     0.55 < m < 0.6 
\label{zone2}
\ee
where the lowest bound is the value of $m$ corresponding to 
${\hat G}=5.88$ from the large-$N$-analysis 
formula (\ref{appelquist}). 
The region (\ref{zone2}) is 
to be compared with (\ref{zone}). As we shall see in section 5, the dynamically
generated mass in our scenario, which does not resort to a 
large-$N$ treatment, lowers significantly the upper bound in (\ref{zone2}).

It is important to emphasize that,
as a consequence of (\ref{awf}),(\ref{photonwff}),  
the non-amputated vertex ${\cal G}(p)$, defined in (\ref{vertexgen}),
approaches a non-zero {\it finite} constant value 
in the infrared $p=0$. 
Notice that this constant behaviour is 
compatible with the Ward identity in the infrared limit $p=0$, 
thereby implying that the information obtained about a non-trivial 
infrared fixed-point structure is gauge invariant, as it should be.

This completes our analysis of including 
the effects of an almost constant $Z(x)\simeq Z(0)$ in the infrared.
The above considerations justify 
retrospectively the omission of $Z$ in 
deriving the differential equation (\ref{emdenmass}). Indeed,   
a constant shift of ${\hat G}$ inside the derivative  
operators in that equation,
which would account for a constant $Z$, 
does not affect its form. 
These results, therefore,  
constitute a highly non-trivial 
check of our approach. 

\setcounter{equation}{0}
\section{Dynamical Derivation of the Fermion Mass Gap}

In the previous section we have assumed 
the presence of 
a finite fermion mass, 
which we have treated effectively as 
an arbitrary parameter of the model.
In this section we turn to the full 
problem, and study the dynamical generation 
of this mass, by deriving it self-consistently
from the corresponding SD mass-gap equation. 

The equation for the gap $\Sigma (p)$ 
is derived from the graphs of fig. \ref{fig1b}, 
which yields:
\be
A(p)\nd{p} + \Sigma (p) = \nd{p} + A(p)e^2 \int \frac{d^3k}{(2\pi)^3}
{\hat G}(k)^2 \gamma _\mu \frac{1}{\nd{k} + M(k)}\gamma ^\mu \frac{1}{(k-p)^2}
\label{massgap}
\ee
where $M(k) \equiv \Sigma (k)/A(k)$ is the mass function, and 
we have pulled out factors of $A(p)$ appropriately so as to be able 
to define an amputated vertex function ${\hat G}(k)$.
The consistency of the approach will be demonstrated below by the 
existence of solutions. 
 
Taking the trace in the above equation, 
and performing the angular integration, 
one arrives easily at:
\bea
&~&M(p) = 3e^2 \int \frac{d^3k}{(2\pi)^3}{\hat G}^2(k) \frac{M(k)}{k^2 + 
M^2(k)}  
\frac{1}{(k-p)^2} =\nn \\
&~&\frac{6\alpha}{\pi^2 p}\int dk \frac{k}{k^2 + M^2(k)}{\hat G}(k)^2 M(k){\rm ln}\left|\frac{k + p}{k - p}\right|
\label{tracem}
\eea
From the middle equation (\ref{tracem}) it becomes clear that 
the quantity 
$g_R (k) \equiv e  {\hat G}(k)$, 
defined in (\ref{runcoupl}),
plays indeed the r\^ole of a ``running coupling''. 
The situation is analogous, but not identical, to that of \cite{am}, 
where a ``running coupling'' has also been obtained from the gap 
equation in the context of a large $N$ analysis.
However, in that case, 
one assumed an Ansatz for the vertex, 
satisfying a truncated form of 
the Ward identities. 
Instead, in our approach there was no necessity 
for a vertex Ansatz, since the non-perturbative
vertex function was determined self-consistently
from the SD equations. As a consequence, 
the running coupling 
(\ref{runcoupl}) is constructed out 
of the amputated vertex function alone; 
the latter is a manifestly
gauge-independent quantity, at least perturbatively.  

Passing into dimensionless variables, in units of $\alpha=e^2/8$, 
${\tilde M}\equiv M(k)/\alpha$, $x\equiv p/\alpha$, $y \equiv k/\alpha$,
and working in the regime of low momenta $x << 1$ as usual, one 
obtains after some straightforward algebra, involving a truncated  
expansion of the logarithmic kernel: 
\be
{\tilde M}(x) =\frac{12}{\pi^2} \left[ 
\frac{1}{x^2}\int _0^x dy \frac{y^2 {\tilde M}(y)}{y^2 + {\tilde M}^2(y)}{\hat G}^2(y) + \int _x^\infty dy \frac{{\tilde M}(y)}{y^2 + {\tilde M}^2(y)}{\hat G}^2(y)\right]
\label{tildemeq}
\ee
Differentiating twice with respect to $x$ one arrives at:
\be
x \frac{d^2}{d x^2} {\tilde M}(x) + 3 \frac{d}{dx}{\tilde M}(x) 
+ \frac{24}{\pi^2}\frac{{\tilde M}(x)}{x^2 + {\tilde M}^2(x)}{\hat G}(x)^2 = 0
\label{gapeq}
\ee
Given that dynamical mass generation is expected to be 
an infrared phenomenon,
we now restrict ourselves to the region~\footnote{Note that this 
is the opposite limit
than the one usually considered in the framework of large-$N$ 
analysis of dynamical mass generation, where one arrives at a 
gap equation by making the assumption $\alpha >> p >> m$
for the momenta $p$ of the pertinent excitations.}, 
\be
x^2 << {\tilde M}^2 << 1
\label{condmx}
\ee
In this region we neglect $x^2$ next to ${\tilde M}^2$ in (\ref{gapeq})
and   
use the solution (\ref{es}) for ${\hat G}(x) \simeq 
{\tilde M}\sqrt{\frac{3}{8}}\pi c$ as $x \rightarrow 0$. The result is:
\be
x \frac{d^2}{d x^2} {\tilde M}(x) + 3 \frac{d}{dx}{\tilde M}(x) 
+ 9c^2{\tilde M}(x)= 0, \qquad x \rightarrow 0
\label{bessel}
\ee
Changing variables $x \rightarrow \xi = x^{-1}$, the equation 
reduces to 
a Bessel equation~\cite{kamke} 
\be
\xi^2 \frac{d^2}{d \xi^2} {\tilde M}(\xi) - \xi \frac{d}{d\xi}{\tilde M}(\xi) 
+ 9c^2 \xi^{-1} {\tilde M}(\xi)= 0, \qquad \xi \rightarrow \infty
\label{bessel2}
\ee
with (formally) general solution:
\be
     {\tilde M}(\xi) = C_1 \xi J_{-2}(-6c \xi^{-1/2}) + C_2 \xi 
Y_{-2}(-6c \xi^{-1/2})        
\label{solutionbess}
\ee
where $C_i,i=1,2$ are arbitrary constants, 
$J_{-n} (x) = (-1)^n J_{n}(x)$, $n = 1,2,3, \dots$, 
is a Bessel function of the first kind, 
and $Y_{n}(x)$ is a generalized  
Bessel function of the second kind~\cite{kamke}. The latter is only defined 
for positive $n$ and this imposes the choice $C_2=0$ in (\ref{solutionbess}). 
Expressing the solution in terms of $x$, $x\rightarrow 0$, 
one has the following power series expression for the dynamical mass:
\be
  {\tilde M}(x) = C_1 x^{-1} \sum _{n=0}^{\infty} (-1)^n  
\left(3c\right)^{2n + 2} x^{1 + n} \frac{1}{ n! \Gamma (3 + n)}
\simeq \frac{9}{2} C_1 c^2 + {\cal O}(x), \qquad x \rightarrow 0
\label{fermionmass}
\ee
From this one obtains the following relation between ${\hat G}(0)$ and 
${\tilde M}(0)\equiv m_f /\alpha $:
\be
       {\tilde M}(0) \equiv m_f/\alpha 
= \left(\frac{12}{\pi^2}\right)^{1/3}C_1^{1/3}{\hat G}^{2/3}(0) 
\label{appelgm}
\ee
This result is to be compared the result (\ref{appelquist})
within the context of a large-$N$ analysis. 
In particular, at first sight it seems that the relation 
(\ref{appelgm}) does not have a critical coupling, above which 
dynamical mass generation occurs. 
However, because the result (\ref{appelgm}) has been derived
in the context of the solution (\ref{es}), 
one should bear in mind the restrictions characterizing that
situation, in particular (\ref{newcond}). 
This implies an appropriate restriction for $C_1$~\footnote{The
restriction (\ref{newcond}) is to be viewed as a boundary condition. 
We remind the reader that the requirement of finiteness of ${\tilde M}$ 
had already fixed the other constant 
$C_2$ to zero.},
\be
{\hat G}^{5/3}(0) - 
\frac{12\sqrt{6}}{5}{\hat G}^{2/3}(0) - 
\frac{8}{3 \left(12 \pi C_1\right)^{1/3}}\left(1 - 
\frac{1}{\pi}\right) < 0 , \qquad 
{\hat G}(0) > \frac{12\sqrt{6}}{5}
\label{c1}
\ee
This restriction implies a critical coupling, 
${\hat G}_c = 12\sqrt{6}/5 \simeq 5.88$ 
but it is derived
in a way independent of any large-$N$ analysis. 
The way to understand (\ref{c1}) is the following: one should first 
fix a range of ${\hat G}(0)$, with ${\hat G}(0) > 5.88$, and then 
use a $C_1$ that will be such that, within this range of the couplings,
Eq. (\ref{c1}) is satisfied for masses ${\tilde m} << 1$. 
As can be readily seen, the bound for $C_1$ obtained from the 
requirement 
that $m <<1$ is far less restrictive than the one 
associated with (\ref{c1}), provided 
${\hat G}(0)$ is not too close to the critical ${\hat G}_c$,
where the mass $m$ vanishes.  
For instance, for ${\hat G}(0) ={\cal O}(8)$, the upper bound 
on $C_1 $ from (\ref{c1}) is of order ${\cal O}(10^{-4})$, while 
for ${\hat G}(0) = 6$ the upper bound is $C_1 < 4$. Notice that 
the bound is 
very sensitive to small changes in ${\hat G}(0)$.    

\begin{figure}[htb]
\epsfxsize=3in
\bigskip
\centerline{\epsffile{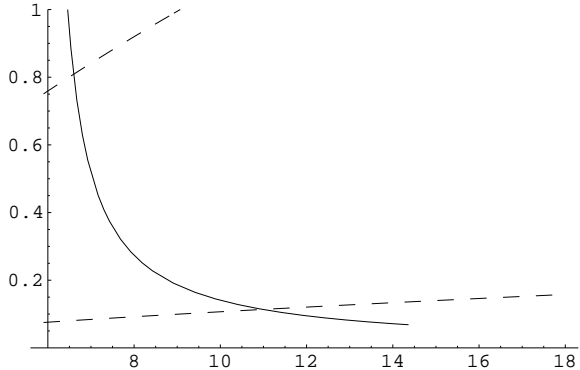}}
\caption{\it\baselineskip=12pt 
Fermion mass versus the infrared-value 
of the coupling ${\hat G}(0)$
using (\ref{appelgm}) (dashed curves), for two values of 
$C_1=10^{-5}$ (lower dashed curve) and $C_1=10^{-2}$ (upper dashed curve). 
The continuous curve is (\ref{fincond}), viewed as a boundary condition.
The value $C_1 =10^{-2}$ should be excluded on grounds of yielding 
too high values of the mass ${\tilde m}$.}  
\bigskip
\label{gmpm}\end{figure}

A typical situation is depicted in fig. \ref{gmpm} for two values 
of $C_1 =10^{-5}, 10^{-2}$. We observe that the case $C_1=10^{-2}$ yields 
an upper bound in the mass which is of order $0.8$ and hence 
should be discarded on the basis that it is not small enough.
On the other hand, the value $C_1=10^{-5}$ yields an acceptable upper bound
$ m \sim 0.1$. In that case, from fig. \ref{gmpm}, we observe that 
the allowed region of $m$ is 
\be
0.08 \lsim m \lsim 0.12
\label{zone4}
\ee
to be compared with (\ref{zone2}), derived using the result 
for the mass in the context of a large-$N$ analysis~\cite{app}. 
The corresponding regime of the couplings ${\hat G}(0)$ 
is:
\be
   5.88 < {\hat G}(0) < 11
\label{regionc4}
\ee

Before closing this section we would like to 
discuss possible applications of the above behaviour; specifically, 
the restriction 
(\ref{regionc4}) of the allowed values of ${\hat G}(0)$ 
may have interesting applications in case
the above model turns out to describe the physics of high-temperature
superconductors. We remind the reader that 
in such models dynamical mass generation coincides with 
superconductivity~\cite{dor}-\cite{fm}. In that case, 
the fermion fields (`electrons') of $QED_3$ 
represent `holons', i.e. electrically-charged
excitations of fermionic statistics,  
which are constituents of the physical electron.
The latter is 
believed to exhibit an effective spin-charge 
separation~\cite{anderson} in the complicated ground-state 
of high-temperature superconductors.
The `photon' of $QED_3$ then represents
an effective Heisenberg spin-spin antiferromagnetic interaction,
responsible for binding the holons in Cooper-like pairs.
In some models~\cite{dor,fm} 
the effective (gauge-invariant) 
coupling ${\hat G}(0)$ may be expressed  
in terms of the parameters 
of the microscopic condensed-matter lattice systems, 
whose long-wavelength
limit is equivalent to the above $QED_3$ model, as:
\be
    {\hat G}(0)^2 \sim \frac{J}{e^2} (1-\eta)
\label{dopheis}
\ee
where 
$\eta$ expresses
the concentration of impurities in the system (doping), and 
$J$ denotes the Heisenberg (antiferromagnetic) exchange energy. 
Hence, on account of 
(\ref{regionc4}), Eq. (\ref{dopheis}) implies 
that $ 6 \lsim (J/e^2)(1 - \eta) \lsim 11$ for superconductivity 
to occur. 
In 
phenomenologically acceptable models~\cite{dor}
$e^2/J \sim 0.1$, which implies 
an upper bound on $\eta \sim 0.4$.
However, the reader should bear 
in mind that the above-described limiting values
are rather indicative at present, given 
that a complete quantitative understanding of the 
underlying dynamics of high-temperature superconductivity
from an effective gauge-theory point of view  
is still lacking. 

\setcounter{equation}{0}
\section{Equation for the Vertex in the case of non-zero photon mass} 

In this section we study the case where a (small) covariant photon 
mass $\delta $ is added to the photon propagator (\ref{photon}):
\be
     \Delta _{\mu\nu}(k) \sim \frac{\delta_{\mu\nu}}{k^2 + \delta ^2 }
\label{photon2}
\ee
For the purposes of our analysis in this 
work, such a mass term will simply be added by hand, 
without further discussions about its origin. 
However, we point out for completeness that 
a small photon mass may be the result 
of non-perturbative configurations (instantons) 
in compact $U(1)$ three-dimensional electrodynamics~\cite{polyakov}. 
Moreover, a term acting like 
a photon mass term arises in real-time finite-temperature
considerations~\cite{aitchison}; in the latter case, of course, one 
loses manifest Lorentz covariance. 

It is straightforward to see that, 
in the presence of a photon 
mass, small compared to the scale 
$\alpha=e^2/8$, 
the resulting integral equation for the amputated vertex reads:
\be
   {\hat G}(p) = \frac{1}{24\pi^2}\frac{e^2}{p}
\int \frac{dk}{k}{\hat G}^3(k)
{\rm ln}\left|\frac{(k + p)^2 +\delta ^2}{(k-p)^2 +  \delta^2}\right| 
\label{vertphot}
\ee
We are interested in the limit $p << \delta $, which would allow us
to study the effects of a photon mass 
on the infrared regime of the theory. 
Expanding the logarithm in (\ref{vertphot}) appropriately, 
we obtain:
\be
    {\hat G}(x) =\frac{4}{3\pi^2}\frac{1}{{\hat \delta} ^2}
\int _0^x dy {\hat G}^3(y) + \frac{4}{3\pi^2}\int _x^\infty 
\frac{dy}{y^2 + {\hat \delta}^2} {\hat G}^3(y) 
\label{eqf}
\ee
where ${\hat \delta} \equiv \delta/\alpha$, $x=p/\alpha$. 

Differentiating once with respect to $x$ one obtains:
\be
     \frac{d}{dx}{\hat G}(x) =\frac{4}{3\pi^2{\hat \delta ^2}}{\hat G}^3 (x) \frac{x^2}{x^2 + {\hat \delta}^2}
\label{deriv}
\ee
which can be integrated to yield
\be
{\hat G}(x) =\frac{1}{\sqrt{c + \frac{8}{3\pi^2{\hat \delta}^2}
\left({\hat \delta }{\rm arctg}(\frac{x}{{\hat \delta}}) - x\right)}}
\label{rcdelta}
\ee
where $c$ is an integration constant to be fixed by the boundary 
condition imposed by the integral equation (see below).

\begin{figure}[htb]
\epsfxsize=3in
\bigskip
\centerline{\epsffile{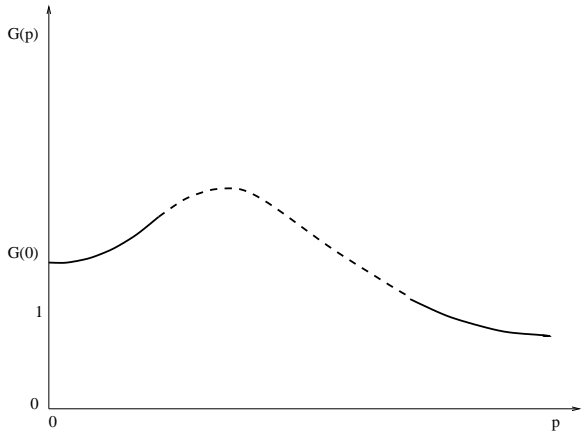}}
\caption{\it\baselineskip=12pt 
Running coupling (versus momentum) 
in the case of a small photon mass.
Note the bump at low momenta, which is absent in the 
case of non-zero fermionic masses. The dashed part of the 
curve is conjectural at present, and can only be derived numerically.
Continuous curves are the result of analytic studies.} 
\bigskip
\label{fig5}\end{figure}

The running coupling $G(x)$ tends to ${\hat G}(0) = \frac{1}{\sqrt{c}}$ as 
$x \rightarrow 0$. 
There is a non-trivial fixed point at $x=0$,
given that the renormalization-group $\beta$-function (\ref{beta}) 
vanishes:
\be 
    x\frac{d}{dx}{\hat G}(x)  \rightarrow  
\frac{4}{3\pi^2}{\hat G}^3 (x) \frac{x^3}{x^2 + {\hat \delta}^2}
=0, \qquad {\rm as}~x \rightarrow 0
\label{fp2}
\ee
Notice that the function ${\hat G}(x)$ increases monotonically 
with increasing $x$, for small $x$.
We next note that by choosing $c$ appropriately one 
can satisfy the condition (\ref{limit})  
${\hat G}(0) >> \sqrt{3/2}$, 
required for self-consistency of the approximation
of ignoring the 
effects of the inhomogeneous term $Z(p)$ in the infrared. 
In such a case 
one expects that at larger $x$ the increase of the coupling stops
at a certain (small) $x=x_0$, and then the 
coupling starts decreasing to reach asymptotically the 
perturbative result ${\hat G} \rightarrow 1$ (see fig. \ref{fig5}). 
The existence of a local maximum is characteristic of the effects
of the photon mass on the infrared behaviour, to be contrasted
with the monotonically decreasing situation in the 
case of a non-zero 
fermion mass (see fig. \ref{fig4}). 
We expect that in the general case, when both photon and
fermion masses are present, the running coupling will exhibit 
a local maximum at low $x$ 
when 
the photon mass is larger than the fermion mass, whilst this 
behaviour  
will be replaced by a monotonic decrease,  
in the case when the fermion mass is considerably larger
than the photon one. This situation should be compared with the 
corresponding one in four-dimensional QCD
in the presence of non-vanishing gluon and 
quark masses~\cite{pc}.

Before closing we discuss briefly a physical situation 
which could be qualitatively similar to 
the case of a covariant infrared cut-off in the 
form 
of a non-zero photon mass, discussed in this section. 
This situation has been argued in \cite{am} 
to simulate finite-temperature effects, given that in such a case
the photon propagator acquires a longitudinal plasmon mass term: 
\be
     \Delta_{\mu\nu}(p_0=0, P \rightarrow 0, T) =
\frac{\delta _{\mu 0}\delta _{0\nu}}{P^2 + M_{00}^2 } + {\rm longitudinal~parts}
\label{photfinitet}
\ee
where, for simplicity, we restricted ourselves to the instantaneous approximation~\cite{dor}, $p_0=0$, and the plasmon mass term 
\be
    M_{00}(T) =\sqrt{\frac{2\alpha {\rm ln}2 T}{\pi}}
\label{moo}
\ee
 We observe from the form (\ref{photfinitet}) that the presence 
of the mass term $M_{00}$ behaves for low $T$ somewhat analogously
(but not identical to) a small photon cut-off mass $\delta$.

\begin{figure}[htb]
\epsfxsize=3in
\bigskip
\centerline{\epsffile{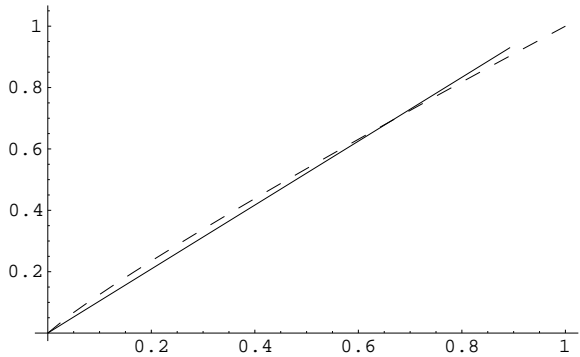}}
\caption{\it\baselineskip=12pt 
Plot of the resistivity (\ref{tbeh}) 
versus temperature $T$ (in units of $\alpha$), for a low-temperature 
region (continuous curve), and comparison with 
the experimentally observed  
single-scaling (basically linear)  
behaviour $T^{1-\chi}, \chi=0.1$ (dashed curve). 
The agreement
is very good for a regime of temperatures accessible to  
experiment.}
\bigskip
\label{fig6}\end{figure}

Prompted by this observation we would like now to
make some speculations 
regarding the low temperature effects of the $\delta $ cut-off term 
in (\ref{rcdelta}). We hasten to emphasize that 
a discussion could at best only  
qualitatively correct, given that a quantitative understanding would require
a finite-temperature extension of the above analysis. Due to the 
loss of Lorentz covariance in the respective propagators the analysis 
is far more complicated than the zero-temperature one.
However, when the temperature $T$ is sufficiently low, its effects on the 
pole structure of the photon propagator
may be simulated by a covariant photon-mass cut-off, at least 
qualitatively~\cite{am,aitchison}. 
For this reason, we are motivated
to examine the effects of ${\hat \delta } <<1$
on the low-temperature scaling of the resistivity~\cite{csm} $\rho$;
the latter
is defined as the inverse of the conductivity 
$\sigma _f$, which expresses the response of the system to 
a change in an externally-applied electromagnetic potential $A_\mu$. 
Ohm's law, in a gauge with $A_0=0$, then, gives~\cite{ioffe,am,csm}:
\be
    \sigma _f \equiv \rho ^{-1} = \frac{1}{\left(P^2 + p_0^2\right) \left(1 + 
\Pi (P,p_0,T) \right)}\left|_{P=0}\right.
\label{condt}
\ee
where $\Pi (P,p_0,T)$ denotes the 
dimensionless photon polarization used in 
this work. Following the analysis of \cite{aitchison} we may assume
for our discussion below that~\footnote{The analysis 
of \cite{aitchison} is based on a real-time
finite temperature approach. In such an approach the 
resistivity is defined in the infrared limit 
$p_0 \rightarrow 0, P=0$ in the presence of a plasmon mass term $M_{00}$
(\ref{moo}) in the photon propagator; as noted in \cite{aitchison},
the way in which the two mass scales ($p_0,P$) approach $0$ 
suffers from ambiguities,
related to physical Landau-damping processes; 
it is not our purpose here to resolve such issues.} :  
\be 
p_0^2/\alpha \rightarrow M_{00}^2/\alpha  \sim {\hat \delta} ^2 \sim 
\frac{2{\rm ln}2}{\pi} T
\label{deltaT}
\ee
From the discussion  following (\ref{runcoupl}) one immediately 
sees that the resistivity, defined for $P=0,x^2={\hat \delta}^2$, 
would be given
in terms of the running coupling ${\hat G}(p)$ by:  
\be
     \rho = {\hat \delta} ^2/{\hat G}^2(0,T) 
\label{rho}
\ee
Using (\ref{rcdelta}) in the limit $x^2={\hat \delta} ^2$, one has:
\be
   \rho \sim \frac{2c T}{\pi}{\rm ln}2 
+ \frac{8}{3\pi^2}\left(\frac{\pi}{2} - 1\right)
\sqrt{\frac{2T}{\pi}{\rm ln}2} 
\label{tbeh}
\ee
The function $\rho$ is plotted in figure (\ref{fig6}), versus a single-power
scaling behaviour of the form: $\rho \sim T^{1-\chi}$, 
with $\chi$ small; the latter is characteristic  
of certain theoretical condensed-matter models~\cite{byczuk},   
and has been used
in order to bound 
possible deviations from
the 
experimentally-observed 
linear-$T$ behaviour
of the resistivity of high-temperature superconductors
at optimal doping~\cite{csm}.

We observe that, upon appropriate choice of the integration constant $c$, 
the temperature dependence of  (\ref{tbeh}) for a low-temperature regime,
accessible to experiment,
is hardly distinguishable from the single power scaling behaviour.

\section{Conclusions and Outlook} 

In this work we have presented a study of chiral-symmetry breaking 
in $QED_3$, based on a system 
of coupled non-linear SD equations.
The novel ingredient is the introduction of 
a semi-amputated vertex, whose dynamics is
governed by a suitably truncated SD equation
of cubic order. This allows for a self-consistent 
determination of the infrared value of the effective charge,
in the presence of infrared cutoffs provided by either 
the fermion mass or a covariant photon mass. 
The theory is characterized by a (gauge-invariant)
non-trivial infrared fixed-point, suggestive of  
non-Fermi liquid behaviour~\cite{shankar,am}. 

The non-linear vertex equation furnishes 
highly non-trivial constraints among the infrared value of 
the effective coupling and the mass of the fermions.
When these constraints are combined 
with the fermion mass-gap equation, 
they select specific regions of the coupling space 
for which dynamical mass generation 
occurs. 
It will be interesting to study 
how these constraints are affected
if one goes beyond the one-loop dressed 
approximation considered here. 

From the physical point of view, we remark that 
dynamical fermion mass generation in $QED_3$ is associated
with superconductivity~\cite{dor,fm}.  
In the relevant 
statistical models, whose low-energy effective theories
are of $QED_3$ type, the couplings depend on the 
doping concentration. 
Therefore the 
restrictions in the parameter space 
we have found above might impose  
restrictions in the allowed models.

An interesting feature of our approach 
is the fact that the infrared structure of the photon 
propagator,
derived as a consistent solution of a 
SD equation, 
implied formally a confining effective potential.
The fact that 
this behaviour occurs in the chiral-symmetry broken phase 
of the theory, appears to be in agreement with generic expectations
that confinement is a sufficient condition for chiral symmetry 
breaking. 

An obvious next step in this programme is the 
attempt to solve the coupled system of SD equations
numerically, without 
the approximation of ignoring the effects of the 
inhomogeneous term $Z(p)$ in the infrared. 
We remind the reader that
in the present work we have restricted 
the allowed values of the coupling such that this
approximation was self-consistent. This, however, 
does not imply 
that solutions with $Z(p)$ not meeting these requirements
are impossible; to establish or exclude their existence 
one should solve the whole 
system of equations, 
which at present can only be done numerically.   
Such a numerical treatment will have the additional 
advantage of not resorting to 
the approximations made to the kernel of the integral equations
in order to convert them into 
differential ones. 

Furthermore, an analysis at finite temperatures 
will reveal whether our conjecture in section 6, on the 
scaling behaviour of the resistivity, is correct.
As we have discussed there, this issue acquires 
great importance in view of very accurate experiments
in high-temperature superconductors indicating a
basically linear scaling with temperatures of the resistivity 
in the normal phase (no fermion mass gap) of the system.

Finally, an extension of these ideas to the non-Abelian 
case appears as a challenge for the future, especially in view
of recent claims~\cite{fm,wen} that non-Abelian gauge symmetries 
might describe the dynamics of 
spin-charge separation in realistic  
antiferromagnetic condensed-matter systems~\cite{anderson}.

\section*{Acknowledgements}

The authors  would like to thank I.J.R. Aitchison for useful 
discussions. 
The work of N.E.M. is partially supported 
by a United Kingdom P.P.A.R.C. Advanced Research Fellowship, 
and that of JP is funded by a Marie Curie Fellowship (TMR-ERBFMBICT 972024).

\end{document}